\documentclass[alpha-refs, twocolumn]{wiley-article}
\usepackage[utf8]{inputenc}

\usepackage{siunitx}
\usepackage{multirow}

\title{Virtual Reality-Based Preoperative Planning for Optimized Trocar Placement in Thoracic Surgery: A Preliminary Study}

\author[1]{Arash Harirpoush}

\author[2]{George Rakovich}

\author[1,3]{Marta Kersten-Oertel}

\author[1,3]{Yiming Xiao}

\affil[1]{Department of Computer Science and Software Engineering, Concordia University, Quebec, Canada}
\affil[2]{Maisonneuve Rosemont Hospital, University of Montreal, Quebec, Canada}
\affil[3]{School of Health, Concordia University, Quebec, Canada}

\begin{document}

\begin{frontmatter}
\maketitle

\begin{abstract}
Video-assisted thoracic surgery (VATS) is a minimally invasive approach for treating early-stage non-small-cell lung cancer. Optimal trocar placement during VATS ensures comprehensive access to the thoracic cavity, provides a panoramic endoscopic view, and prevents instrument crowding. While established principles such as the Baseball Diamond Principle (BDP) and Triangle Target Principle (TTP) exist, surgeons mainly rely on experience and patient-specific anatomy for trocar placement, potentially leading to sub-optimal surgical plans that increase operative time and fatigue.
To address this, we present the first virtual reality (VR)-based pre-operative planning tool with tailored data visualization and interaction designs for efficient and optimal VATS trocar placement, following the established surgical principles and consultation with an experienced surgeon. In our preliminary study, we demonstrate the system's application in right upper lung lobectomy, a common thoracic procedure typically using three trocars.  A preliminary user study of our system indicates it is efficient, robust, and user-friendly for planning optimal trocar placement, with a great promise for clinical application while offering potentially valuable insights for the development of other surgical VR systems.

\keywords{Virtual reality, thoracic surgery planning, image-guided-surgery,  video‐assisted thoracic surgery}
\end{abstract}
\end{frontmatter}

\section{Introduction}
Lung cancer is the second most common cancer and the leading cause of cancer-related deaths worldwide\citep{sung2021global}. In the United States, approximately 56,000 to 57,000 lung cancer resections are performed each year, with lobectomies being the most common type of resection \citep{SHC8191}. Low post-trauma minimally invasive surgeries, such as video-assisted thoracoscopic surgery (VATS), are now being used to treat early-stage non-small-cell lung cancer \citep{bendixen2016postoperative}. During VATS surgeries, optimal trocar placement, which guides the entry of surgical tools and endoscopic camera into the body through small incisions is necessary for surgical success. Optimal placement involves three key principles: (1) Trocars must be carefully positioned to ensure full access to all relevant areas within the thoracic cavity to facilitate complete surgical exploration and intervention. (2) The endoscopic camera trocar should be strategically placed to provide a panoramic view of the surgical field and sufficient room for instrument manipulation and avoiding visual obstruction. (3) All trocar placements should be meticulously planned to prevent instrument crowding or ``fencing", ensuring smooth and efficient instrument handling throughout the procedure \citep{sasaki2005triangle, landreneau1992video}.

While VATS offers numerous benefits, the optimal placement of trocars remains an area of limited research and standardized guidelines. Two common principles to guide trocar placement exist: (1) the Baseball Diamond Principle (BDP), which offers enhanced maneuverability and wider access to the thoracic cavity, particularly advantageous in non-pulmonary procedures \cite{ismail2014comparing}, and (2) the Triangle Target Principle (TTP), which optimizes direct access to the surgical target and is preferred for retraction or stapling \citep{sasaki2005triangle, ismail2014comparing}. Despite these principles, surgeons primarily rely on their experience and patient-specific anatomy to make trocar placement decisions \cite{sasaki2005triangle}, potentially leading to longer operating times, increased risk of complications, and greater fatigue for the surgical team due to limited instrument working area, and maneuverability \citep{preda2020preoperative}. Thus there is a need for effective preoperative planning techniques, such as through virtual reality (VR) for precise and effective trocar placement.

In this paper, we introduce the first VR application for thoracic pre-operative planning to efficiently provide optimal trocar placement based on established surgical principles and developed in close collaboration with an experienced thoracic surgeon. In a preliminary study, we showcase the system's application in right upper lung lobectomy, a common thoracic surgery. Following conventional practice, we included three trocars: two for surgical instruments in tissue resection and manipulation and one for the insertion of an endoscopic camera for surgical monitoring. The importance of accessing all areas of the chest cavity in this procedure led to the development of a rule-based trocar placement system. This system aims to help in precise trocar placement to optimize the operable area, i.e., the intersection between the working area of surgical instruments and the endoscopic camera's field of view (FOV).

We designed three key VR interaction and visualization features that are tailored for thoracic surgery. \textbf{First}, to enhance precision in planning, our application uses a pivot mechanism for surgical tool trocar placement. \textbf{Second}, we employed a "hand grabbing" interaction method for endoscopic camera position planning and camera trocar placement. \textbf{Lastly}, real-time visual feedback and evaluation metrics were devised to further assist in trocar placement based on existing guidelines and discussions with an experienced thoracic surgeon. Upon completion of planning, a comprehensive summary is generated, detailing key metrics for surgical plan quality to allow further refinement of plans. A preliminary user study was done to confirm the system's robustness and usability. The resulting insights can provide valuable information for future development of VR surgical applications for thoracic procedures and beyond.

\section{Related Works}
\subsection{Patient-specific 3D Models}
Recent studies have highlighted the significant advantages of incorporating patient-specific 3D models into preoperative planning across various surgical specialties \citep {cen2021three, bakhuis2023essential, ujiie2021developing, preda2020preoperative}. Within thoracic surgery, \cite{cen2021three} demonstrated the utility of both physical (3D printed) and digital (VR/MR) 3D models in improving surgical field alignment during complex pulmonary atresia surgeries \citep {cen2021three}. \cite{ujiie2021developing} focused on lung segmentectomy, utilizing a VR-based system with patient-specific 3D lung models to enhance surgical planning and surgeon confidence by facilitating the identification of anatomical landmarks and potential surgical challenges.

The value of 3D models extends beyond thoracic procedures. In laparoscopic hiatal hernia repair, \cite{preda2020preoperative} developed a preoperative planning system based on patient-specific 3D reconstruction and simulation, receiving positive feedback from surgeons who noted its potential to improve ergonomics and its particular value in challenging cases involving obese patients with large hiatal hernias.
Further evidence for the utility of 3D models in thoracic surgery comes from \cite{bakhuis2023essential}, who compared 2D planning with CT images to 3D planning in VR for pulmonary segmentectomy. Their findings revealed that 2D planns were adjusted in 52\% of cases and tumor localization was inaccurate in 14\%, underscoring the potential of 3D models to improve surgical accuracy and planning \citep {bakhuis2023essential}. Beyond their use in individual procedures, \cite{heuts2016preoperative} explored the broader benefits of 3D models in thoracic surgical planning, finding that their use increases surgical efficiency, minimizes complications, and enhances overall surgical outcomes \citep {heuts2016preoperative}. 

\subsection{Extended Reality Applications in Minimally Invasive Surgeries}

Extended Reality (ER) has been used in various minimally invasive surgeries to enhance procedural efficiency and precision. Several studies \citep{lopez2013design, feuerstein2005automatic, feuerstein2008intraoperative} have explored the use of ER for trocar planning systems to optimize minimally invasive surgery outcomes. For instance, \cite{lopez2013design} developed an augmented reality (AR) system to improve trocar placement accuracy in laparoscopic cholecystectomy, which is facilitated by a full HD monitor with transparency for enhanced depth perception. In their study involving four clinicians and 24 patients, the AR system demonstrated an 63\% improvement in accuracy compared to traditional trocar placement methods.
Similarly, \cite{feuerstein2005automatic} presented an AR system for port placement in robotic-assisted surgeries (RATS). Their approach involved registering the patient for their preoperative CT scan by maneuvering the endoscope around fiducials, enabling automatic 3D position reconstruction. Later, \cite{feuerstein2008intraoperative} proposed an AR system for port placement and intraoperative planning in minimally invasive liver resection that further accounts for intraoperative organ shifts. In another study, \cite{bauernschmitt2007optimal} reported a significant reduction in operation time in minimally invasive robot‐assisted heart surgery, thanks to employing their AR system for offline port placement planning and intraoperative navigation.
Meanwhile, other endeavors \citep{simoes2013leonardo, schwenderling2022augmented} have proposed decision-based  mixed‐reality (MR) and AR systems for automatic path planning to enhance surgical performance and streamline surgical workflows. For example, \cite{simoes2013leonardo} introduced a decision-aid MR system to improve RATS performance and reduce planning time. Their system incorporates an optimization algorithm that suggests trocar placements based on the patient's anatomy and the specific surgery type. These suggestions are then projected onto the patient's body with a projector, allowing surgeons to refine the placement as needed.
In another study, \cite{schwenderling2022augmented} proposed a condition-based automated path planning AR system for percutaneous interventions. This system uses a projector to visualize the insertion point, path quality, and target on a phantom. Their results demonstrated the potential of visualizing insertion points and path quality in selecting safer access paths.

Beyond surgical planning, virtual Reality (VR) environments with haptic feedback devices have emerged as valuable tools for simulating surgical procedures and training trocar placement. Addressing limitations in previous training modules, such as limited anatomical variation, \cite{ solomon2011simulating} proposed a VR training system with haptic feedback to simulate VATS right upper lobectomy. In their system, trocar placement for each instrument is selected from predetermined sites on the chest wall, and instruments are then controlled via haptic devices. The process begins with determining the 30-degree thoracoscope trocar location, followed by an inspection of the anatomy through a camera view to guide the placement of the remaining trocars. The system includes both training and testing modes, with the latter featuring pop-up questions and explanations for incorrect answers.
Similarly, \cite{haidari2022simulation} developed a VR system with haptic devices for simulating VATS resection of the five lung lobes. Their study involved surgeons across three experience levels: novice, intermediate, and experienced. Their results showed significant differences between novices and experienced surgeons in blood loss, procedure time, and total instrument path length. Meanwhile, the only significant difference between intermediates and experienced surgeons was in procedure time.

While previous studies have widely investigated the influence of ER environments and patient-specific 3D models in surgical planning, the use of HMD VR systems for trocar placement in VATS remains untouched. This method could enhance surgical outcomes by offering surgeons superior depth perception and spatial understanding compared to traditional AR-based or monitor-based methods. Furthermore, using a VR environment could decrease potential registration errors that may arise in AR systems, thereby contributing to increased precision in surgical planning.

\section{Materials and Methods}
\subsection{3D Model Generation}
A 3D thoracic anatomical model was constructed based on a patient computed tomography (CT) scan (1.5$\times$1.5$\times$1.5 $mm^{3}$ resolution) selected from the publicly available TotalSegmentator~\cite{wasserthal2023totalsegmentator} dataset. We obtained anatomical segmentations of the vertebrae, ribs, scapula, and trachea, which were manually refined using 3D Slicer to enhance model accuracy. Additionally, we further manually segmented the pulmonary vasculature and skin surface with 3D Slicer. All segmentations were converted into triangulated meshes (.obj format), and then integrated into the VR environment.

\begin{figure*}[h!]
    \centering
    \includegraphics[width=\textwidth]
    {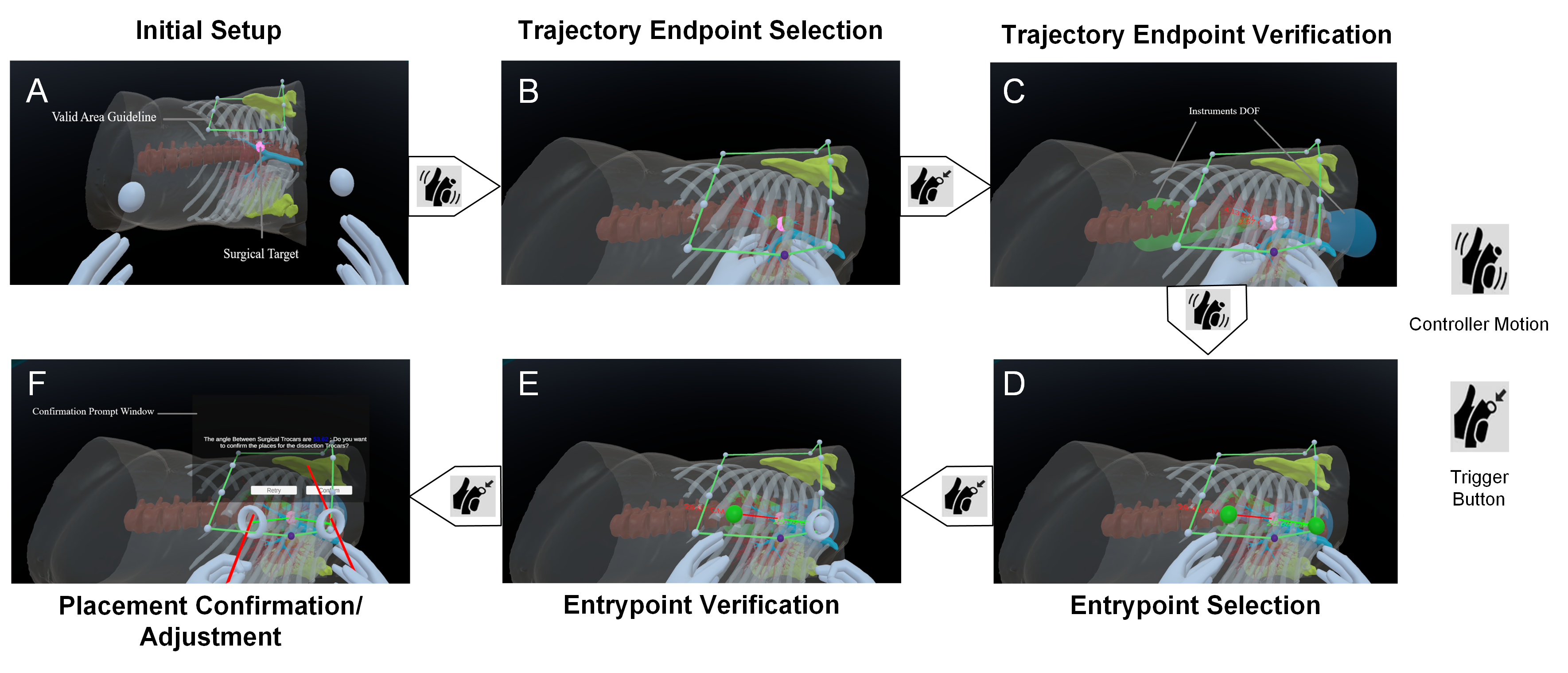}
    \caption{
    Overview of the pivot mechanism in surgical trocar placement:
    A. Initial anterior view with trajectory endpoint spheres positioned in front of each controller; B. Spheres manipulated to define endpoints (green when near target); C. Endpoint verification displays working area and trajectory paths; D. Spheres moved to the skin to define entry points (green on contact); E. Green spheres and paths indicate valid entry, verifying trocar placement; F. Manipulation angle displayed for adjustment/confirmation.
    }
    \label{fig:surgicalTrocarPlacement}
\end{figure*}

\subsection{VR user interface and workflow} \label{WorkFlowExplanation}
Our system was created using the Oculus Quest Pro headset and controllers, employing the Unity game engine (Version 2021.3.11f1). Both development and user studies were conducted on a desktop computer with an NVIDIA GeForce RTX 3090 GPU, an 11th Gen Intel® Core™ i9 CPU, and 32 GB of RAM.
The VR environment developed for this study includes three main visual components.  \textbf{First}, a large information panel is positioned in front of the user to provide instructions for surgical planning tasks. \textbf{Second}, a virtual screen is positioned to the right of the information panel to display simulated video streaming from the virtual endoscopic camera, enabling precise adjustments and optimal positioning of the camera. \textbf{Third}, a detailed 3D anatomical model, featuring distinctly color-coded anatomical structures (see Fig. 1A, vertebrae in brown, scapula in yellow, trachea in blue, and pulmonary vasculature in red) is placed in front of the user for surgical planning. In the 3D model, we annotated the convergent point of the surgical tool trajectories and the optical axis of the endoscopic camera as a pink sphere. This convergent point was identified by our collaborating surgeon as the
root of the right upper lobe and is common for planning most lung procedures. As key anatomies in surgical planning, we render the skin and ribs as semi-transparent structures to allow views of the underlying anatomy and their spatial relationship.

The workflow of the system is as follows. During the surgical planning, the user will remain in a standing position, mimicking a surgeon's posture during surgery. Before initiating planning, the user is asked to re-adjust the vertical position of the anatomical model to a comfortable level by using a slider selection tool shown in a control panel in the VR environment. Afterwards, planning can be initiated by pressing the ``Start" button on the control panel.
Typically during the right upper lung lobectomy procedure, the surgeon operates from the front of the patient (anterior view) while the camera-holding assistant is positioned at the back (posterior view). Therefore, the positioning of the patient model will be automatically adjusted according to this convention for the two surgical planning tasks in sequence: (1) surgical tool trocar placement with an anterior view of the patient, replicating the surgeon's perspective, and (2) endoscopic camera and the associated trocar placement, with a posterior view that mirrors the assistant's perspective. This task sequence was refined through an iterative development process to enhance workflow efficiency. In both tasks, the system provides visual feedback as color cues and numerical metric displacement in VR to guide users toward valid trocar placement areas. Further details on the data visualization and interaction schemes are provided in Sections \ref{toolPlacement} and \ref{cameraPlacement}.

\subsubsection{Surgical tool trocar placement} \label{toolPlacement}
The trocar placement uses a pivot mechanism guided by two white spheres, one attached at the tip of each controller (Fig. \ref{fig:surgicalTrocarPlacement}A). This mechanism consists of two phases: \textit{endpoint selection }and \textit{entry point placement} for the surgical trajectories. \textbf{First}, the user reaches the two white spheres from left and right controllers within a 3D anatomical model towards the convergent point (the pink sphere) until the sphere turns green (Fig. \ref{fig:surgicalTrocarPlacement}B) indicating correct endpoint localization. The endpoints (i.e., white spheres) are placed by pressing the corresponding controller's trigger button. Afterward, a red surgical trajectory line will extend from the placed endpoint to each controller, along with a 20-degree-angle cone, the angle between the side to the principal axis, that represents the degree-of-free (DOF) of the surgical instrument's maneuver. The cone angle was defined using the surgeon's wrist range of motion (40 degrees for radial-ulnar deviation), as indicated by previous research \citep{RYU1991409}. Note that the right trocar's DOF cone is indicated by green color and the left one's by blue (Fig. \ref{fig:surgicalTrocarPlacement}C). 

\textbf{Second}, the user drags the trajectory lines with the controllers onto the skin surface while ensuring that they avoid bony structures and that the real-time displayed trajectory distance for each controller remains under 28 cm, which is the maximum working length of the surgical instruments. The user must place the trocars in the designated area as contoured by green lines on the anatomical model. When these criteria are met, the system provides visual cues by turning both the trajectory lines and spheres green (Fig. \ref{fig:surgicalTrocarPlacement}D). The user then fixes the placement of each of the two trocars by pressing the corresponding controller's trigger button (Fig. \ref{fig:surgicalTrocarPlacement}E). After fixing both trocars, the ``manipulation angle" between the two trajectories is displayed on a confirmation panel to confirm the planning or repeat the procedure till satisfaction (Fig. \ref{fig:surgicalTrocarPlacement}F).
Note that prior research \citep{hanna1997optimal} suggests a manipulation angle between 45 and 75 degrees for optimal surgical instrument positioning with trocars parallel and sufficiently spaced. 

\subsubsection{Endoscopic camera placement} \label{cameraPlacement}
\begin{figure*}[ht!]
    \centering
    \includegraphics[width=0.9\textwidth]
    {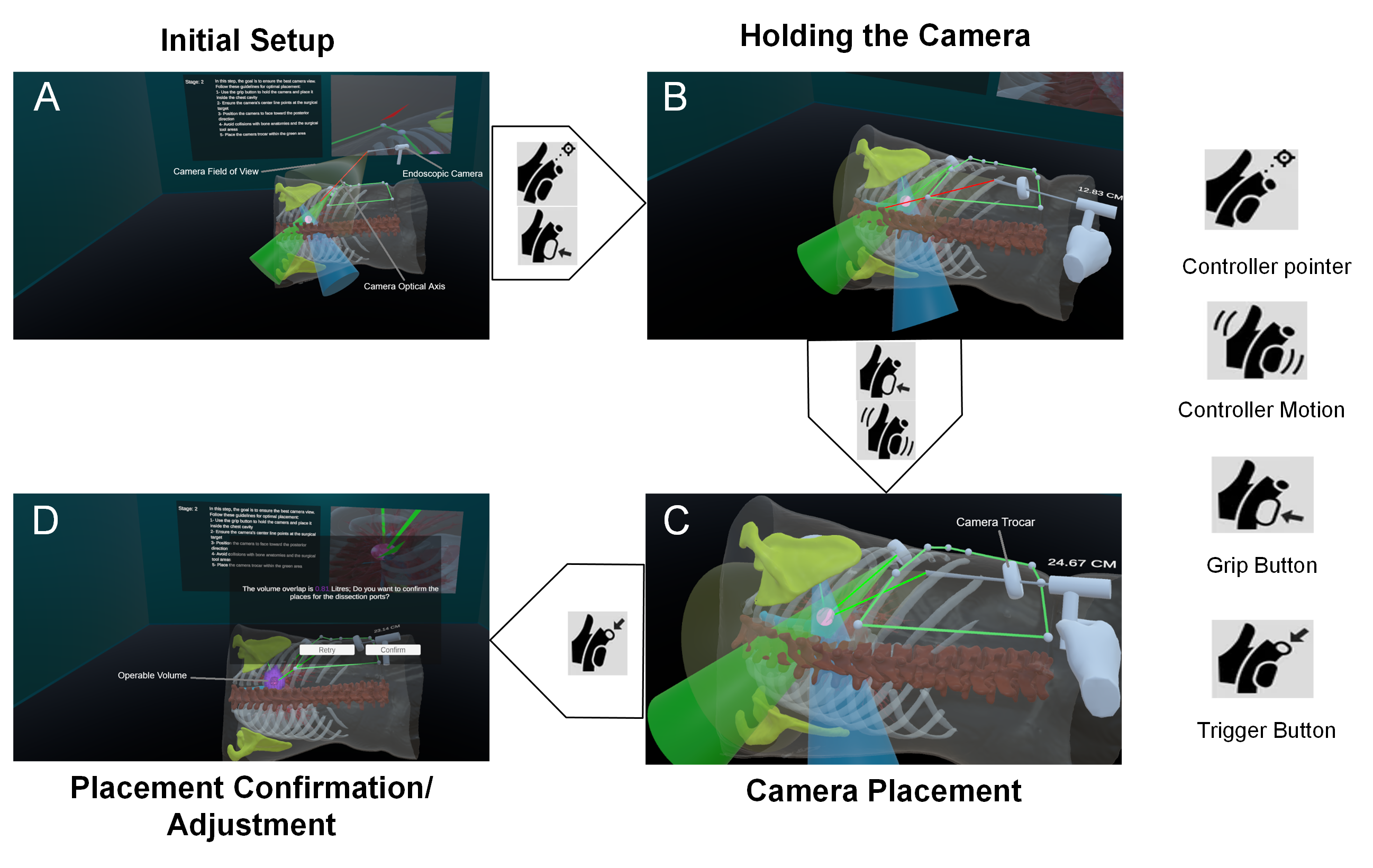}
    \caption{
    Overview of the hand grabbing method in camera placement:
    A. Initial posterior view and endoscopic camera; B. Pointing toward endoscopic camera and hold it by pressing grip button; C. Green camera optical axis line demonstrates valid placement; D. Volume of operable area displayed for adjustment/confirmation.
    }
    \label{fig:CameraPlacement}
\end{figure*}

For our system, we simulate a rigid endoscopic camera (an elongated tube with the camera at the tip) with a 30-degree tilt angle (between the optical axis and the rigid tubular body of the camera) and a 60-degree field of view, which is preferred for thoracic surgery \citep{luh2006video}. During the task of endoscopic camera placement, we visualize the camera's FOV as a semi-transparent yellow cone and the optical axis as a red line (Fig. \ref{fig:CameraPlacement}A). The user can manipulate the camera using a hand-grabbing interaction, by pressing the grip button of their dominant controller to hold and release it to place it in space(Fig. \ref{fig:CameraPlacement}B). The second task of surgical planning requires the user to insert the camera into the chest cavity, by aiming the optical axis towards the convergent point (pink sphere) and checking the virtual camera display for optimal views. Upon inserting the camera tube into the body, a virtual trocar appears intersecting the skin surface, marking the camera's entry point and guiding the user to position it within the designated area (as contoured by green lines on the anatomical model).
To avoid instrument crowding, the camera should be positioned outside the working area (shown as blue and green cones) of the surgical tools. Further, contact with bony structures should be avoided. To ensure correct placement, the red line (camera optical axis) will turn green once it aims directly at the convergent point without obstructions (Fig. \ref{fig:CameraPlacement}C). Upon pressing the trigger button of the controller, a confirmation panel will appear to confirm or repeat the placement. Upon confirmation, the operable volume that considers the surgical tools' DOFs and camera's FOV will be calculated and visualized as purple voxels with numerical quantification in liters (Fig. \ref{fig:CameraPlacement}D). 

\subsection{Computing operable volume}
For the surgery, it is desirable to maximize the area that both surgical tools can cooperate while the endoscopic camera can inspect the full operation of the tools. Thus, the operable volume is determined by the overlap between the surgical tools' DOF and the camera's FOV, represented as three different cones. While triangulated meshes accurately represent the surface of objects, they do not provide the volume of the mesh. To address this, we employed the mesh voxelization method introduced by \citep{WolfireVoxelization} to compute the operable volume. This consists of three steps: 
(1) A 3D grid surrounding the given mesh will be created, forming the foundation for the process with each cell representing a voxel.
 (2) The mesh surface will be voxelized by identifying voxels intersecting with the mesh triangles, effectively replacing the triangulated representation with small 3D cubes.
 (3) A scan-line fill algorithm will be used to identify the voxels within the object border. This process is similar to filling a shape in 2D by drawing horizontal lines until the boundaries are reached.
To balance accuracy and efficiency, we use $1.5 cm \times 1.5 cm \times 1.5 cm$ voxels; smaller voxels would improve resolution but increase computational cost. We customized the implementation of \cite{mattatzUnityVoxel}, which was based on the work of \citep{WolfireVoxelization} to compute the operable volume. Specifically, to compute the volumetric overlap between multiple meshes, we use a single 3D grid covering all models. Each mesh is assigned a unique ID (one for each cone), and for each voxel, the mesh ID is stored in a HashSet. Overlapping voxels are identified by HashSets containing the same number of elements as the input meshes.

\subsection{User study design \& system validation}
Upon informed consent, we recruited 20 non-clinician participants (age = 25.95 $\pm$ 3.31 years, 7 female, 13 male) for our user study. To better understand the study cohort, we also surveyed their level of familiarity with VR technology and human anatomy. Among them, 75\% indicated "Familiar" or "Somewhat Familiar" with VR, while only 30\% reported similar familiarity with human anatomy, with one participant indicating "Unfamiliar" with both. All participants were right-handed, and two (one male, one female) reported color blindness. No participants experienced VR sickness. 

Participants were first given a brief Powerpoint presentation introducing the clinical context, tasks, and goals of the study. Following this, a hands-on tutorial was conducted to familiarize participants with the VR environment, planning process, and various interactions. This tutorial involved tasks different from those in the main study. During the tutorial, participants practiced planning on the left side of the 3D patient model, with an anterior view provided. Text-to-speech technology for the instruction from the information panel was integrated to offer assistance throughout each task. For the camera placement task, a semi-transparent "phantom camera" positioned at the desirable location and position was presented as a ground truth reference, and the participants were asked to place the actual camera to overlap with the phantom guide. This served to illustrate optimal camera placement and angling towards the posterior side of the patient, as required in the surgery. Participants were encouraged to continue practicing until they felt comfortable using the system. Following the tutorial, we conducted the user study to formally validate our proposed system by following the workflow introduced in Section \ref{WorkFlowExplanation}.

The proposed system was evaluated through a mixed-methods approach employing both semi-quantitative and quantitative measures. System usability was assessed using the System Usability Scale (SUS) by \cite{brooke1996sus}, a widely recognized standardized questionnaire. The SUS evaluation is a Likert-scale questionnaire consisting of ten items, each with a range of 1 (strongly disagree) to 5 (strongly agree) \citep{lewis2018system}. Questions alternate between positively and negatively worded statements, ensuring participants actively engage with the content and thoughtfully consider their responses. These questions cover various aspects of the system, including effectiveness, efficiency, and overall user satisfaction. Among the 10 questions of SUS, each odd-numbered question is scored as x-1, and each even-numbered question is scored as 5-x, where x is the question’s resulting value. The scores for each participant are then summed, and then multiplied by 2.5 - resulting in a maximum SUS score of 100. A software system that receives an SUS score above 68 indicates good usability. 

To further evaluate participant experience and effectiveness of the tailored data visualization and interaction designs, an additional Likert-scale questionnaire with eleven items was used to assess engagement, immersion, system usability, and the efficacy of visualizations, interactions, and visual feedback (the questions are detailed in Fig. \ref{fig:UXQuestions}). Specifically, the participants were asked to evaluate their engagement level within the application, the application's visual appeal, and usefulness in the designated task as well as the ergonomic design of the system.  They were also asked to evaluate the ease of use and effectiveness of specific functionalities, including pivoting methods for surgical trocar placement,  the hand-grabbing for camera placement, the visual feedback mechanisms provided, the information panels, and the final visualization of the operable volume. Participants rated each item on a 1-to-5 Likert scale (1=strongly disagree, 5=strongly agree). Finally, participants were asked to provide open-ended feedback on the positive and negative aspects of the system, along with recommendations for system improvement, and reported their familiarity with virtual reality (VR) and human anatomy. For the total SUS score, a one-sample t-test was used to assess whether the results were significantly different from 68. For each SUS sub-score and the customized UX questions, we compared the results to a neutral response (score=3), also with the Mann–Whitney U test. A $p-value<0.05$ was used to indicate a statistically significant difference. 

In addition to the semi-quantitative assessment, relevant quantitative metrics were collected from the proposed VR system for each designated task. These metrics included the total time spent on each task, the number of adjustments made in each task, and the historical and final positions of the trocars and the camera. For the first task (surgical trocar placement), we also recorded trajectory distance (in cm) for each surgical instrument (measured as the distance between the skin entry point and the surgical target), as well as the manipulation angle (the angle between the instruments upon reaching the surgical target). For the second task (camera placement), the volume of overlap between the camera's field of view and the surgical instruments' working area (in liters) was recorded.

\section{Results}
\subsection{Semi-Quantitative Evaluation}
\begin{figure*}[h!]
    \centering
    \includegraphics[width=\textwidth]
    {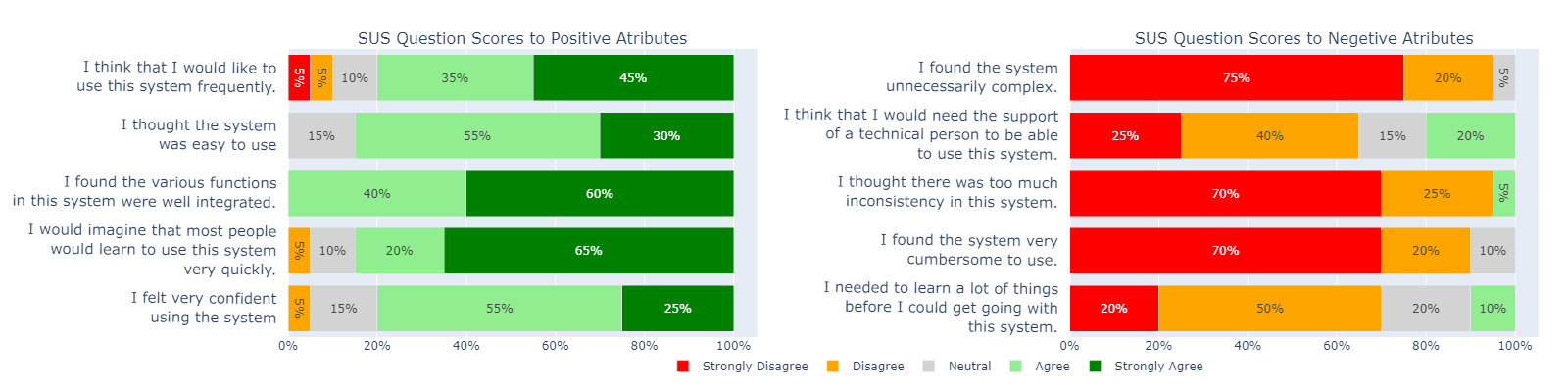}
    \caption{
    Distribution of SUS Question Scores Across Participants.
    }
    \label{fig:SUSScores}
\end{figure*}

Our VR system achieved an average SUS score of $81.8\pm10.5$, significantly higher than the usability threshold of 68 (p=$1.24\times10^{-5}$), categorizing it as ``A" in system usability \citep{brooke1996sus}. In addition, all scores of individual SUS and user experience (UX) questions are significantly better than the neutral score of 3 (p<0.001). The distributions of individual SUS question scores are illustrated in Fig. \ref{fig:SUSScores}. These results indicate positive experience and attitude for various aspects of the proposed system. Specifically, the SUS questionnaire responses highlighted that participants perceived the system as well-integrated (score = $4.6\pm0.5$) but expressed lower confidence in task performance (score = $4.0\pm0.8$). While they did not find the system complex (score = $1.3\pm0.6$), they indicated a preference for technical support (score = $2.3\pm1.1$).

\begin{figure*}
    \centering
    \includegraphics[width=\textwidth]
    {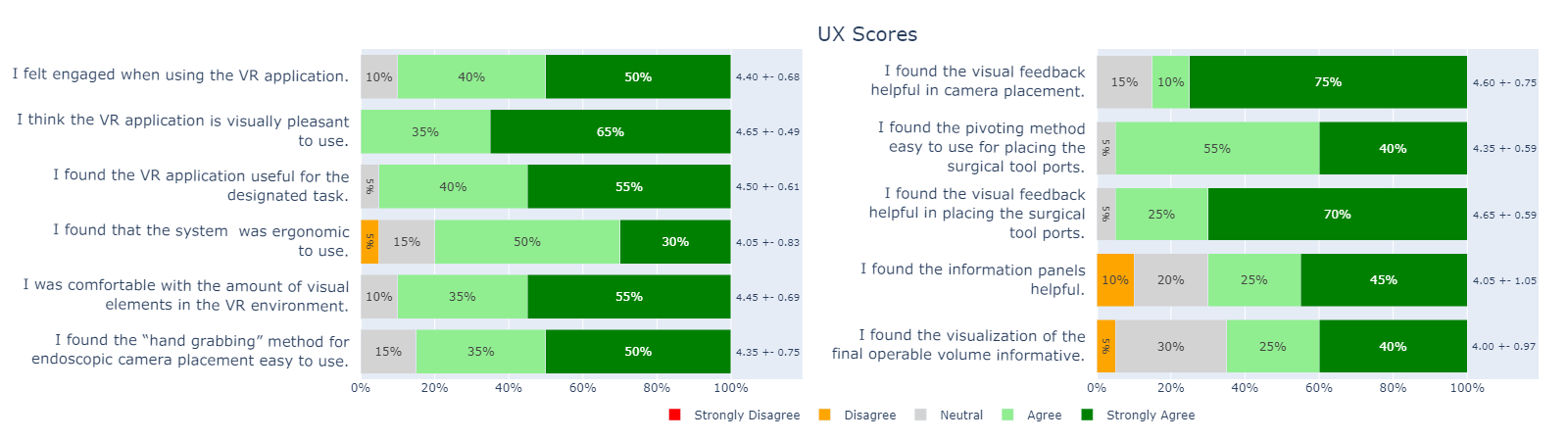}
    \caption{
    Distribution of UX Question Scores across Participants, with mean $\pm$ standard deviation displayed beside the respective bar plot.
    }
    \label{fig:UXQuestions}
\end{figure*}

For the UX questions, all average ratings ranged from 4 to 4.65, with a majority of respondents expressing positive feedback (rating 4 or 5 out of 5) on various aspects. Specifically, 65\% found the final visualization informative, 80\% found the system ergonomic, 90\% felt engaged, and 85\% found the hand-grabbing interface and visual feedback for camera placement intuitive. The majority of participants (95\%) also found the pivot method for trocar placement intuitive, while 70\% found the information panels helpful. The assessments of the individual UX questions are depicted in Fig. \ref{fig:UXQuestions}.

In the open-ended questions, 19 out of 20 participants provided positive and negative aspects of the surgical planning system. Most (14/19) found it easy to use and the feedback metrics helpful (7/19). However, two participants noted the semi-transparent materials hindered depth perception, though visual feedback (White spheres turn into green) helped. Nine participants suggested improvements, e.g., four recommended auditory feedback for guidance and errors, four suggested more guidance for how to optimize surgical planning, such as color-coded manipulation angles on the confirmation panel, and one participant proposed direct 3D model manipulation for height adjustment of the 3D model.

\subsection{Quantitative Evaluation}
Trajectory distance, manipulation angle, operable volume, and task completion times were collected from the VR application. In Task 1 (surgical trocar placement), the maximum trajectory distance for both trocars was less than 28 cm, ensuring the surgical target was reachable. The average manipulation angle of 48 degrees was consistent with recommendations from prior research. For Task 2 (camera placement), positioning the camera outside the DOF of other trocars prevented instrument interference and maximized the common area volume, averaging 1 liter of operable volume across participants. We also recorded the number of adjustments and time required for each task during the user study. The summary of these data can be seen in Table \ref{tab:QuantitativeResults}.

The majority of participants (75\%) completed both tasks without adjustments. Participants spent an average of $3.70\pm1.52$ minutes on planning, with Task 1 taking  $1.37\pm0.89$ minutes and Task 2 taking $2.33\pm1.15$ minutes. Our statistical analysis also revealed significant negative correlations between time spent on the surgical planning and anatomy familiarity (p = 0.041 and correlation = -0.460). This suggests familiarity with the human anatomy can boost performance efficiency.
\begin{table}[htbp]
\renewcommand{\arraystretch}{1.5} 
\caption{Quantitative Evaluation from the User Study} 
\label{tab:QuantitativeResults}
\resizebox{0.5\textwidth}{!}{ 
\begin{tabular}{|l|l|l|} 
\hline
\multirow{2}{*}{\textbf{Task}} & \centering \multirow{2}{*}{\textbf{Metric}} & \multirow{2}{*}{\textbf{Result}} \\
& &  \\
\hline
\multirow{5}{*}{Surgical Trocar placement} & Time (Minutes) & 1.37 $\pm$ 0.89 \\

     & Number of Adjustments & 0.35 $\pm$ 0.67 \\
    
     & Manipulation Angle & 48.63 $\pm$ 7.39 \\
    
     & Right Hand Trajectory Distance (CM) & 25.13 $\pm$ 1.84 \\
    
     & Left Hand Trajectory Distance (CM) & 27.07 $\pm$ 0.70 \\
    \hline
    \multirow{3}{*}{Camera Placement} & Time (Minutes) & 2.33 $\pm$ 1.15 \\
    
    & Number of Adjustments & 0.35 $\pm$ 0.67 \\
    
     & Volume of Common Workable Area (Litres) & 1.01 $\pm$ 0.12 \\
\hline
\end{tabular}
}
\end{table}

\section{Discussion}
In an earlier version of our system, mirroring standard thoracic surgical procedures, participants were required to position the endoscope camera before placing surgical trocars. However, a pilot study involving four participants revealed the necessity for camera adjustments after trocar placement to mitigate instrument crowding and optimize the shared workspace. Consultation with our expert surgeon led to the decision to reverse the task order in the final system. Although in typical surgical procedures, the camera is placed before surgical trocars to guide following placements, employing semi-transparent materials in our 3D model enables the view of internal anatomies in our system making this sequence unnecessary. By reversing the task order, we eliminated the redundant camera adjustment step and the potential for instrument fighting during camera placement.

In the semi-quantitative evaluation using the SUS questionnaire and customized UX questions showed promising results. While participants generally found the system well-integrated and easy to use, a lack of confidence and a perceived need for technical support emerged. This may be related to the absence of a definitive metric for optimal surgical view and manipulation angles, despite the incorporation of soft metrics to guide trocar placement. The UX questions highlighted a positive user experience overall, with high engagement and perceived usefulness, which are crucial for future clinical adoption. However, information panels and the final operable volume visualization were slightly less well-received than other items in the UX questions. Participants suggested a voice assistant for guidance and error reporting. Notably, those who found the operable volume visualization informative reported lower system complexity and less need for technical support in the SUS questionnaire, resulting in higher overall SUS scores. Regarding the ``freehand" camera placement and pivot mechanism, most participants responded favorably and found the visual feedback helpful. Notably, 15\% of participants held a neutral view of the freehand camera placement and its feedback, compared to only 5\% for surgical trocar placement, suggesting an area for potential improvement. Finally, with a short planning time ($3.70\pm1.52$ minutes) with no failed surgical plans, our proposed system offers high efficiency and robustness, required for clinical use.

The current study has several limitations. First, semi-transparent rendering of anatomical structures (e.g., ribs, skin) compromised depth perception. Second, varying difficulty levels for trocar placement based on surgical target location and individual anatomy were not fully explored due to time constraints and the use of one patient model. Third, the limited number of anatomical structures included in the 3D model, due to visualization challenges and computational complexity of segmentation, restricted the development of comprehensive metrics of the proposed system. For example, incorporating the chest wall muscles could help in defining metrics to avoid thick muscles in the chest wall, which can minimize tissue damage and bleeding, while maximizing ease of motion during camera placement. Finally, in our preliminary study, we only recruited non-clinicians for system validation due to the limited accessibility to thoracic surgeons, although the system development greatly benefited from the expertise of our surgical collaborator. Future work will focus on addressing these limitations through alternative visualization techniques, a wider range of patient models, refining the system's metrics and guidelines in collaboration with clinicians, and additional clinical participants in extended system validation upon further refinement.

\section{Conclusion}
In this paper, we present the first pre-operative planning VR system designed to optimize trocar placement in thoracic lung surgeries. Our system incorporates an effective pivoting mechanism and a hand-grabbing method, both seamlessly integrated with visual feedback, to help users in the planning process. A comprehensive user study revealed promising results regarding system usability and overall user satisfaction. The insights from the VR system design and assessment can provide important information for similar surgical VR system development, which has a profound potential in clinical practice.

\bibliography{refs}

\begin{thebibliography}{27}
\expandafter\ifx\csname natexlab\endcsname\relax\def\natexlab#1{#1}\fi
\expandafter\ifx\csname url\endcsname\relax
  \def\url#1{\texttt{#1}}\fi
\expandafter\ifx\csname urlprefix\endcsname\relax\def\urlprefix{URL: }\fi

\bibitem[{Bakhuis et~al.(2023)Bakhuis, Sadeghi, Moes, Maat, Siregar, Bogers and Mahtab}]{bakhuis2023essential}
Bakhuis, W., Sadeghi, A.~H., Moes, I., Maat, A.~P., Siregar, S., Bogers, A.~J. and Mahtab, E.~A. (2023) Essential surgical plan modifications after virtual reality planning in 50 consecutive segmentectomies.
\newblock \textit{The Annals of Thoracic Surgery}, \textbf{115}, 1247--1255.

\bibitem[{Bauernschmitt et~al.(2007)Bauernschmitt, Feuerstein, Traub, Schirmbeck, Klinker and Lange}]{bauernschmitt2007optimal}
Bauernschmitt, R., Feuerstein, M., Traub, J., Schirmbeck, E.~U., Klinker, G. and Lange, R. (2007) Optimal port placement and enhanced guidance in robotically assisted cardiac surgery.
\newblock \textit{Surgical endoscopy}, \textbf{21}, 684--687.

\bibitem[{Bendixen et~al.(2016)Bendixen, J{\o}rgensen, Kronborg, Andersen and Licht}]{bendixen2016postoperative}
Bendixen, M., J{\o}rgensen, O.~D., Kronborg, C., Andersen, C. and Licht, P.~B. (2016) Postoperative pain and quality of life after lobectomy via video-assisted thoracoscopic surgery or anterolateral thoracotomy for early stage lung cancer: a randomised controlled trial.
\newblock \textit{The Lancet Oncology}, \textbf{17}, 836--844.

\bibitem[{Brooke et~al.(1996)}]{brooke1996sus}
Brooke, J. et~al. (1996) Sus-a quick and dirty usability scale.
\newblock \textit{Usability evaluation in industry}, \textbf{189}, 4--7.

\bibitem[{Cen et~al.(2021)Cen, Liufu, Wen, Qiu, Liu, Chen, Yuan, Huang and Zhuang}]{cen2021three}
Cen, J., Liufu, R., Wen, S., Qiu, H., Liu, X., Chen, X., Yuan, H., Huang, M. and Zhuang, J. (2021) Three-dimensional printing, virtual reality and mixed reality for pulmonary atresia: early surgical outcomes evaluation.
\newblock \textit{Heart, Lung and Circulation}, \textbf{30}, 296--302.

\bibitem[{Feuerstein et~al.(2008)Feuerstein, Mussack, Heining and Navab}]{feuerstein2008intraoperative}
Feuerstein, M., Mussack, T., Heining, S.~M. and Navab, N. (2008) Intraoperative laparoscope augmentation for port placement and resection planning in minimally invasive liver resection.
\newblock \textit{IEEE Transactions on Medical Imaging}, \textbf{27}, 355--369.

\bibitem[{Feuerstein et~al.(2005)Feuerstein, Wildhirt, Bauernschmitt and Navab}]{feuerstein2005automatic}
Feuerstein, M., Wildhirt, S.~M., Bauernschmitt, R. and Navab, N. (2005) Automatic patient registration for port placement in minimally invasixe endoscopic surgery.
\newblock In \textit{Medical Image Computing and Computer-Assisted Intervention--MICCAI 2005: 8th International Conference, Palm Springs, CA, USA, October 26-29, 2005, Proceedings, Part II 8}, 287--294. Springer.

\bibitem[{Games(2009)}]{WolfireVoxelization}
Games, W. (2009) Triangle mesh voxelization.
\newblock \urlprefix\url{http://blog.wolfire.com/2009/11/Triangle-mesh-voxelization}.

\bibitem[{Haidari et~al.(2022)Haidari, Bjerrum, Hansen, Konge and Petersen}]{haidari2022simulation}
Haidari, T.~A., Bjerrum, F., Hansen, H.~J., Konge, L. and Petersen, R.~H. (2022) Simulation-based vats resection of the five lung lobes: a technical skills test.
\newblock \textit{Surgical Endoscopy}, 1--9.

\bibitem[{Hanna et~al.(1997)Hanna, Shimi and Cuschieri}]{hanna1997optimal}
Hanna, G., Shimi, S. and Cuschieri, A. (1997) Optimal port locations for endoscopic intracorporeal knotting.
\newblock \textit{Surgical Endoscop}, \textbf{11}, 397--401.

\bibitem[{Heuts et~al.(2016)Heuts, Nia and Maessen}]{heuts2016preoperative}
Heuts, S., Nia, P.~S. and Maessen, J.~G. (2016) Preoperative planning of thoracic surgery with use of three-dimensional reconstruction, rapid prototyping, simulation and virtual navigation.
\newblock \textit{Journal of Visualized Surgery}, \textbf{2}.

\bibitem[{Ismail and Mishra(2014)}]{ismail2014comparing}
Ismail, A.~J. and Mishra, R. (2014) Comparing task performance and comfort during nonpulmo nary video-assisted thoracic surgery procedures between the application of the ‘baseball diamond’and the ‘triangle target’principles of port placement in swine models.
\newblock \textit{World}, \textbf{7}, 60--65.

\bibitem[{Landreneau et~al.(1992)Landreneau, Mack, Hazelrigg, Dowling, Acuff, Magee and Ferson}]{landreneau1992video}
Landreneau, R.~J., Mack, M.~J., Hazelrigg, S.~R., Dowling, R.~D., Acuff, T.~E., Magee, M.~J. and Ferson, P.~F. (1992) Video-assisted thoracic surgery: basic technical concepts and intercostal approach strategies.
\newblock \textit{The Annals of thoracic surgery}, \textbf{54}, 800--807.

\bibitem[{Lewis(2018)}]{lewis2018system}
Lewis, J.~R. (2018) The system usability scale: past, present, and future.
\newblock \textit{International Journal of Human--Computer Interaction}, \textbf{34}, 577--590.

\bibitem[{L{\'o}pez-Mir et~al.(2013)L{\'o}pez-Mir, Naranjo, Fuertes, Alca{\~n}iz, Bueno and Pareja}]{lopez2013design}
L{\'o}pez-Mir, F., Naranjo, V., Fuertes, J., Alca{\~n}iz, M., Bueno, J. and Pareja, E. (2013) Design and validation of an augmented reality system for laparoscopic surgery in a real environment.
\newblock \textit{BioMed Research International}, \textbf{2013}, 758491.

\bibitem[{Luh and Liu(2006)}]{luh2006video}
Luh, S.-p. and Liu, H.-p. (2006) Video-assisted thoracic surgery—the past, present status and the future.
\newblock \textit{Journal of Zhejiang University Science B}, \textbf{7}, 118--128.

\bibitem[{Mattatz(2019)}]{mattatzUnityVoxel}
Mattatz (2019) Unity voxel.
\newblock \url{https://github.com/mattatz/unity-voxel}.

\bibitem[{Potter et~al.(2023)Potter, Puttaraju, Sulit, Beqari, Andrews, Kumar, Sharma, Sharma, Spencer and Yang}]{SHC8191}
Potter, A.~L., Puttaraju, T., Sulit, J.~C., Beqari, J., Andrews, C. A.~M., Kumar, A., Sharma, M., Sharma, M., Spencer, P.~J. and Yang, C.-F.~J. (2023) Assessing the number of annual lung cancer resections performed in the united states.
\newblock \textit{Shanghai Chest}, \textbf{7}.
\newblock \urlprefix\url{https://shc.amegroups.org/article/view/8191}.

\bibitem[{Preda et~al.(2020)Preda, Ciob{\^\i}rc{\u{a}}, Gruionu, Iacob, Sapalidis, Gruionu, Castravete, P{\u{a}}trașcu and \c{S}urlin}]{preda2020preoperative}
Preda, S.~D., Ciob{\^\i}rc{\u{a}}, C., Gruionu, G., Iacob, A.~c., Sapalidis, K., Gruionu, L.~G., Castravete, c., P{\u{a}}trașcu, c. and \c{S}urlin, V. (2020) Preoperative computer-assisted laparoscopy planning for the minimally invasive surgical repair of hiatal hernia.
\newblock \textit{Diagnostics}, \textbf{10}, 621.

\bibitem[{Ryu et~al.(1991)Ryu, Cooney, Askew, An and Chao}]{RYU1991409}
Ryu, J., Cooney, W.~P., Askew, L.~J., An, K.-N. and Chao, E.~Y. (1991) Functional ranges of motion of the wrist joint.
\newblock \textit{The Journal of Hand Surgery}, \textbf{16}, 409--419.
\newblock \urlprefix\url{https://www.sciencedirect.com/science/article/pii/036350239190006W}.

\bibitem[{Sasaki et~al.(2005)Sasaki, Hirai, Kawabe, Uesaka, Morioka, Ihaya and Tanaka}]{sasaki2005triangle}
Sasaki, M., Hirai, S., Kawabe, M., Uesaka, T., Morioka, K., Ihaya, A. and Tanaka, K. (2005) Triangle target principle for the placement of trocars during video-assisted thoracic surgery.
\newblock \textit{European journal of cardio-thoracic surgery}, \textbf{27}, 307--312.

\bibitem[{Schwenderling et~al.(2022)Schwenderling, Heinrich and Hansen}]{schwenderling2022augmented}
Schwenderling, L., Heinrich, F. and Hansen, C. (2022) Augmented reality visualization of automated path planning for percutaneous interventions: a phantom study.
\newblock \textit{International Journal of Computer Assisted Radiology and Surgery}, \textbf{17}, 2071--2079.

\bibitem[{Simoes and Cao(2013)}]{simoes2013leonardo}
Simoes, M. and Cao, C.~G. (2013) Leonardo: A first step towards an interactive decision aid for port-placement in robotic surgery.
\newblock In \textit{2013 IEEE international conference on systems, man, and cybernetics}, 491--496. IEEE.

\bibitem[{Solomon et~al.(2011)Solomon, Bizekis, Dellis, Donington, Oliker, Balsam, Zervos, Galloway, Pass and Grossi}]{solomon2011simulating}
Solomon, B., Bizekis, C., Dellis, S.~L., Donington, J.~S., Oliker, A., Balsam, L.~B., Zervos, M., Galloway, A.~C., Pass, H. and Grossi, E.~A. (2011) Simulating video-assisted thoracoscopic lobectomy: a virtual reality cognitive task simulation.
\newblock \textit{The Journal of thoracic and cardiovascular surgery}, \textbf{141}, 249--255.

\bibitem[{Sung et~al.(2021)Sung, Ferlay, Siegel, Laversanne, Soerjomataram, Jemal and Bray}]{sung2021global}
Sung, H., Ferlay, J., Siegel, R.~L., Laversanne, M., Soerjomataram, I., Jemal, A. and Bray, F. (2021) Global cancer statistics 2020: Globocan estimates of incidence and mortality worldwide for 36 cancers in 185 countries.
\newblock \textit{CA: a cancer journal for clinicians}, \textbf{71}, 209--249.

\bibitem[{Ujiie et~al.(2021)Ujiie, Yamaguchi, Gregor, Chan, Kato, Hida, Kaga, Wakasa, Eitel, Clapp et~al.}]{ujiie2021developing}
Ujiie, H., Yamaguchi, A., Gregor, A., Chan, H., Kato, T., Hida, Y., Kaga, K., Wakasa, S., Eitel, C., Clapp, T.~R. et~al. (2021) Developing a virtual reality simulation system for preoperative planning of thoracoscopic thoracic surgery.
\newblock \textit{Journal of Thoracic Disease}, \textbf{13}, 778.

\bibitem[{Wasserthal et~al.(2023)Wasserthal, Breit, Meyer, Pradella, Hinck, Sauter, Heye, Boll, Cyriac, Yang et~al.}]{wasserthal2023totalsegmentator}
Wasserthal, J., Breit, H.-C., Meyer, M.~T., Pradella, M., Hinck, D., Sauter, A.~W., Heye, T., Boll, D.~T., Cyriac, J., Yang, S. et~al. (2023) Totalsegmentator: Robust segmentation of 104 anatomic structures in ct images.
\newblock \textit{Radiology: Artificial Intelligence}, \textbf{5}.

\end{thebibliography}

\end{document}